\documentclass[12pt]{article}
\usepackage{lineno}
\usepackage{graphicx}
\usepackage{amsmath}
\usepackage{hhline}
\usepackage{amssymb}
\usepackage{times}
\usepackage{cite}
\usepackage{array}
\usepackage{multirow}

\usepackage{color}
\definecolor{rltred}{rgb}{0.75,0,0}
\definecolor{rltgreen}{rgb}{0,0.5,0}
\definecolor{rltblue}{rgb}{0,0,0.75}

\newif\ifpdf
\ifx\pdfoutput\undefined
    \pdffalse          
\else
    \pdfoutput=1       
    \pdftrue
\fi

\ifpdf
\usepackage{thumbpdf}
\usepackage[pdftex,
        colorlinks=false,
        urlcolor=rltblue,       
        filecolor=rltblue,     
        linkcolor=rltblue,       
        pdftitle={Example File for pdflatex},
        pdfauthor={H1},
        pdfsubject={Template file},
        pdfkeywords={High-Energy Physics, Particle Physics},
        pdfpagemode=None,
        bookmarksopen=true]{hyperref}
 
\fi

\newlength{\dinwidth}
\newlength{\dinmargin}
\begin{document}

\newcommand{\pom}{{I\!\!P}}
\newcommand{\reg}{{I\!\!R}}
\newcommand{\slowpi}{\pi_{\mathit{slow}}}
\newcommand{\fiidiii}{F_2^{D(3)}}
\newcommand{\fiidiiiarg}{\fiidiii\,(\beta,\,Q^2,\,x)}
\newcommand{\n}{1.19\pm 0.06 (stat.) \pm0.07 (syst.)}
\newcommand{\nz}{1.30\pm 0.08 (stat.)^{+0.08}_{-0.14} (syst.)}
\newcommand{\fiidiiiful}{F_2^{D(4)}\,(\beta,\,Q^2,\,x,\,t)}
\newcommand{\fiipom}{\tilde F_2^D}
\newcommand{\ALPHA}{1.10\pm0.03 (stat.) \pm0.04 (syst.)}
\newcommand{\ALPHAZ}{1.15\pm0.04 (stat.)^{+0.04}_{-0.07} (syst.)}
\newcommand{\fiipomarg}{\fiipom\,(\beta,\,Q^2)}
\newcommand{\pomflux}{f_{\pom / p}}
\newcommand{\nxpom}{1.19\pm 0.06 (stat.) \pm0.07 (syst.)}
\newcommand {\gapprox}
   {\raisebox{-0.7ex}{$\stackrel {\textstyle>}{\sim}$}}
\newcommand {\lapprox}
   {\raisebox{-0.7ex}{$\stackrel {\textstyle<}{\sim}$}}
\def\gsim{\,\lower.25ex\hbox{$\scriptstyle\sim$}\kern-1.30ex%
\raise 0.55ex\hbox{$\scriptstyle >$}\,}
\def\lsim{\,\lower.25ex\hbox{$\scriptstyle\sim$}\kern-1.30ex%
\raise 0.55ex\hbox{$\scriptstyle <$}\,}
\newcommand{\pomfluxarg}{f_{\pom / p}\,(x_\pom)}
\newcommand{\dsf}{\mbox{$F_2^{D(3)}$}}
\newcommand{\dsfva}{\mbox{$F_2^{D(3)}(\beta,Q^2,x_{I\!\!P})$}}
\newcommand{\dsfvb}{\mbox{$F_2^{D(3)}(\beta,Q^2,x)$}}
\newcommand{\dsfpom}{$F_2^{I\!\!P}$}
\newcommand{\gap}{\stackrel{>}{\sim}}
\newcommand{\lap}{\stackrel{<}{\sim}}
\newcommand{\fem}{$F_2^{em}$}
\newcommand{\tsnmp}{$\tilde{\sigma}_{NC}(e^{\mp})$}
\newcommand{\tsnm}{$\tilde{\sigma}_{NC}(e^-)$}
\newcommand{\tsnp}{$\tilde{\sigma}_{NC}(e^+)$}
\newcommand{\st}{$\star$}
\newcommand{\sst}{$\star \star$}
\newcommand{\ssst}{$\star \star \star$}
\newcommand{\sssst}{$\star \star \star \star$}
\newcommand{\tw}{\theta_W}
\newcommand{\sw}{\sin{\theta_W}}
\newcommand{\cw}{\cos{\theta_W}}
\newcommand{\sww}{\sin^2{\theta_W}}
\newcommand{\cww}{\cos^2{\theta_W}}
\newcommand{\trm}{m_{\perp}}
\newcommand{\trp}{p_{\perp}}
\newcommand{\trmm}{m_{\perp}^2}
\newcommand{\trpp}{p_{\perp}^2}
\newcommand{\alp}{\alpha_s}
\newcommand{\cm}{\,\mbox{cm}}

\newcommand{\alps}{\alpha_s}
\newcommand{\sqrts}{$\sqrt{s}$}
\newcommand{\LO}{$O(\alpha_s^0)$}
\newcommand{\Oa}{$O(\alpha_s)$}
\newcommand{\Oaa}{$O(\alpha_s^2)$}
\newcommand{\PT}{P_{\perp}}
\newcommand{\ptmiss}{\mbox{$\not{\hspace{-0.1cm}\PT}$}\hspace{0.075cm}}
\newcommand{\rpv}{$\not{\hspace{-0.1cm}R_{P}}$}
\newcommand{\Rp}{\mbox{$\not \hspace{-0.15cm} R_p$}}
\newcommand{\JPSI}{J/\psi}
\newcommand{\sh}{\hat{s}}
\newcommand{\uh}{\hat{u}}
\newcommand{\MP}{m_{J/\psi}}
\newcommand{\PO}{I\!\!P}
\newcommand{\xbj}{x}
\newcommand{\xpom}{x_{\PO}}
\newcommand{\ttbs}{\char'134}
\newcommand{\xpomlo}{3\times10^{-4}}  
\newcommand{\xpomup}{0.05}  
\newcommand{\dgr}{^\circ}
\newcommand{\pbarnt}{\,\mbox{{\rm pb$^{-1}$}}}
\newcommand{\gev}{\,\mbox{GeV}}
\newcommand{\WBoson}{\mbox{$W$}}
\newcommand{\fbarn}{\,\mbox{{\rm fb}}}
\newcommand{\fbarnt}{\,\mbox{{\rm fb$^{-1}$}}}
%
%
\newcommand{\qsq}{\ensuremath{Q^2} }
\newcommand{\gevsq}{\ensuremath{\mathrm{GeV}^2} }
\newcommand{\et}{\ensuremath{E_t^*} }
\newcommand{\rap}{\ensuremath{\eta^*} }
\newcommand{\gp}{\ensuremath{\gamma^*}p }
\newcommand{\dsiget}{\ensuremath{{\rm d}\sigma_{ep}/{\rm d}E_t^*} }
\newcommand{\dsigrap}{\ensuremath{{\rm d}\sigma_{ep}/{\rm d}\eta^*} }
\def\Journal#1#2#3#4{{#1} {\bf #2} (#3) #4}
\def\NCA{\em Nuovo Cimento}
\def\NIM{\em Nucl. Instrum. Methods}
\def\NIMA{{\em Nucl. Instrum. Methods} {\bf A}}
\def\NPB{{\em Nucl. Phys.}   {\bf B}}
\def\PLB{{\em Phys. Lett.}   {\bf B}}
\def\PRL{\em Phys. Rev. Lett.}
\def\PRD{{\em Phys. Rev.}    {\bf D}}
\def\ZPC{{\em Z. Phys.}      {\bf C}}
\def\EJC{{\em Eur. Phys. J.} {\bf C}}
\def\CPC{\em Comp. Phys. Commun.}

\begin{titlepage}

%
  
\noindent
DESY 04--084 \hfill ISSN 04818--9833 \\
May 2004

\vspace*{2cm}

\begin{center}
\begin{Large}

{\bf Search for bosonic stop decays in R--parity violating supersymmetry in {\boldmath $e^+ p$} collisions at HERA}

\vspace{2cm}

H1 Collaboration

\end{Large}
\end{center}

\vspace{2cm}

\begin{abstract} \noindent
A search for scalar top quarks in R--parity violating supersymmetry is performed in $e^+ p$ collisions at HERA using the H1 detector. 
The data, taken at \mbox{$\sqrt{s}=319\gev$} and \mbox{$301\gev$}, correspond to an integrated luminosity of $106\pbarnt$.
The resonant production of scalar top quarks $\tilde{t}$ in positron quark fusion via an R--parity violating Yukawa coupling $\lambda'$ is considered with the subsequent bosonic stop decay $\tilde{t}\rightarrow \tilde{b} W$.
The R--parity violating decay of the sbottom quark $\tilde{b}\rightarrow d \bar\nu_e$ and leptonic and hadronic $W$ decays are considered.
No evidence for stop production is found in the search for bosonic stop decays nor in a search for the direct R--parity violating decay $\tilde{t}\rightarrow eq$.
Mass dependent limits on $\lambda'$ are obtained in the framework of the Minimal Supersymmetric Standard Model.
Stop quarks with masses up to $275\gev$ can be excluded at the $95\%$ confidence level for a Yukawa coupling of electromagnetic strength.
\end{abstract}

\vspace{1.5cm}

\begin{center}
To be submitted to Phys. Lett. B
\end{center}

\end{titlepage}

\begin{flushleft}

A.~Aktas$^{10}$,               
V.~Andreev$^{26}$,             
T.~Anthonis$^{4}$,             
A.~Asmone$^{33}$,              
A.~Babaev$^{25}$,              
S.~Backovic$^{37}$,            
J.~B\"ahr$^{37}$,              
P.~Baranov$^{26}$,             
E.~Barrelet$^{30}$,            
W.~Bartel$^{10}$
% ...

\bigskip{\it
% ...
$ ^{4}$ Inter-University Institute for High Energies ULB-VUB, Brussels;
          Universiteit Antwerpen, Antwerpen; Belgium$^{ c}$ \\
% ...
$ ^{10}$ DESY, Hamburg, Germany \\
% ...
$ ^{25}$ Institute for Theoretical and Experimental Physics,
          Moscow, Russia$^{ l}$ \\
 $ ^{26}$ Lebedev Physical Institute, Moscow, Russia$^{ e}$ \\
% ...
$ ^{30}$ LPNHE, Universit\'{e}s Paris VI and VII, IN2P3-CNRS,
          Paris, France \\
% ...
$ ^{33}$ Dipartimento di Fisica Universit\`a di Roma Tre
          and INFN Roma~3, Roma, Italy \\
% ...
$ ^{37}$ DESY, Zeuthen, Germany \\
% ...
$ ^b$ Supported by the UK Particle Physics and Astronomy Research
      Council, and formerly by the UK Science and Engineering Research
      Council \\
 $ ^c$ Supported by FNRS-FWO-Vlaanderen, IISN-IIKW and IWT
      and  by Interuniversity Attraction Poles Programme,
      Belgian Science Policy \\
% ...
$ ^e$ Supported by the Deutsche Forschungsgemeinschaft \\
% ...
$ ^l$ Partially Supported by Russian Foundation
      for Basic Research, grant    no. 00-15-96584 \\
}

\end{flushleft}

\newpage

\section{Introduction}
\noindent
Deep inelastic collisions at HERA are ideally suited to the search for new particles coupling to an electron\footnote{In the following, the term {\it electron} refers to both electrons and positrons.}--quark pair. 
Such particles include squarks ($\tilde q$), the scalar supersymmetric (SUSY) partners of quarks, in models with R--parity violation (\Rp) \cite{SUSY}.
In most scenarios the squarks of the third generation, stop ($\tilde t$) and sbottom ($\tilde b$), are the lightest squarks.
In the present analysis we focus on resonant stop quark production in $eq$--fusion which
proceeds via an \Rp\hspace{0.075cm} coupling $\lambda'$. 
We investigate SUSY scenarios in which the sbottom mass is smaller than the stop mass, \mbox{$M_{\tilde b} < M_{\tilde t}$}, which are complementary to previous \Rp\hspace{0.075cm} SUSY searches for squark production by H1 \mbox{\cite{haller,rpvsusy}}.
This study is particularly interesting following the observation of events with isolated electrons or muons and missing transverse momentum \cite{Andreev:2003pm}.
The dominant Standard Model (SM) source for such events is the production of real $W$ bosons.
Some of these events have a hadronic final state with large transverse momentum and are not typical of SM $W$ production.
These striking events may indicate a production mechanism involving processes beyond the Standard Model, such as the production of a scalar top quark and its decays as proposed in \cite{stop}.

\begin{figure}[p]
  \begin{center}
    \includegraphics[width=0.9\textwidth]{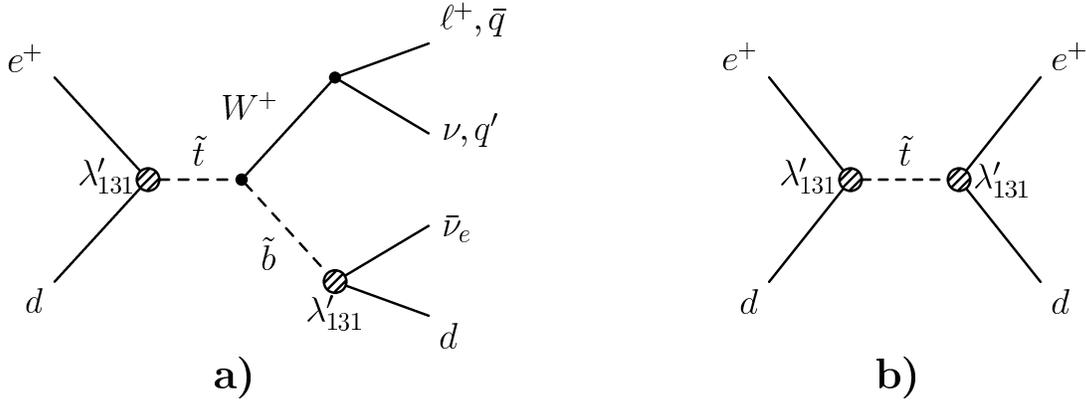}
  \end{center}
  \caption{Lowest order s channel diagram for \Rp\hspace{0.075cm} stop production at HERA followed by a) the bosonic decay of the stop and b) the \Rp\hspace{0.075cm} decay of the stop.}
  \label{fig:feynman}
\end{figure}

In this paper a search is presented for stop quarks which are produced resonantly, \mbox{$e^+q \stackrel{\lambda'}\rightarrow \tilde t$.}
Of particular interest is the bosonic decay $\tilde t \rightarrow \tilde b W$, where the sbottom decay into SM particles, $\tilde b \stackrel{\lambda'}\rightarrow \bar\nu_e d$, is also R--parity violating.
This decay mode is experimentally investigated for the first time. 
The analysis includes both leptonic and hadronic $W$ decays. 
A scenario is investigated, in which decays of the light squarks into neutralinos and charginos are kinematically not possible.
In order to cover all decay modes, the \Rp\hspace{0.075cm} decay $\tilde t \stackrel{\lambda'}\rightarrow e^+ d$ is also considered. 
The corresponding diagrams are shown in figure~\ref{fig:feynman}.
At HERA, stop quarks with masses close to the kinematic limit of $\sim 300\gev$ can be produced.
Such high masses are kinematically inaccessible at LEP and the  bosonic stop decay modes considered are difficult to observe at the Tevatron. 

The analysis uses the data collected with the H1 detector in positron--proton scattering in the years 1994--2000, where the energy of the incoming positron is $E_e = 27.6\gev$.
The proton energy in 1994--1997 is $E_p = 820\gev$, which leads to a centre--of--mass energy of \mbox{$\sqrt{s} = 301\gev$}.
The data  correspond to an integrated luminosity of $\mathcal{L}_{301} = 37.9\pbarnt$. 
In the years 1999 and 2000, where the proton energy is $E_p = 920\gev$ and
the centre--of--mass energy is $\sqrt{s} = 319\gev$, the data recorded
correspond to an integrated luminosity of \mbox{$\mathcal{L}_{319} = 67.9\pbarnt$}.

\section{Phenomenology}
\label{chap:phen}

The most general supersymmetric theory which is gauge invariant with respect to the Standard Model gauge group allows Yukawa couplings between two SM fermions and a squark or a slepton.
These couplings induce violation of R--parity, defined as $R_{p} = (-1)^{3n_B+n_L+2S}$, where $n_B$ is the baryon number, $n_L$ is the lepton number and $S$ is the spin of a particle. 
At HERA the Yukawa couplings between a lepton--quark pair and a squark
are of particular interest \cite{susyth}.
Here the resonant
production of stop quarks and the \Rp\hspace{0.075cm} decay of stop
and sbottom quarks via a non-vanishing coupling $\lambda'_{131}$ are
investigated.
Both processes are described by the Lagrangian 
\begin{equation}
  \mathcal{L}_{\mbox{\Rp}} \sim \lambda'_{131} e_L \tilde{t}_L \bar{d}_R + \lambda'_{131} \nu_{e,L} \tilde{b}_L \bar{d}_R,
\label{eq:lagrangian}
\end{equation}
where the indices $L$ and $R$ denote the left and right states of the fermionic fields and their corresponding scalar superpartners. 
The coupling $\lambda'_{131}$ is a free parameter of the model with the subscripts $131$ being the generation indices.

In the third generation large mixings between $\tilde{q}_L$ and $\tilde{q}_R$ are conceivable \cite{SUSY}.
Because of the structure of the squark mass matrices the stop and sbottom are the most likely candidates for the lightest squark states.
The phases $\theta_{\tilde{q}}$ (with $\tilde{q}=\tilde{t}$ or $\tilde{q}=\tilde{b}$) parameterise the mass eigenstates,
\begin{eqnarray}
    \tilde{q}_1 = \tilde{q}_L \cos\theta_{\tilde{q}} + \tilde{q}_R\sin\theta_{\tilde{q}}\;\;\; \mbox{and}\;\;\;
    \tilde{q}_2 = -\tilde{q}_L\sin\theta_{\tilde{q}} + \tilde{q}_R\cos\theta_{\tilde{q}},
\end{eqnarray}
with the convention $M_{\tilde{q}_1}< M_{\tilde{q}_2}$.
Since the \Rp\hspace{0.075cm} stop interaction involves only the $\tilde{t}_L$ component of the fields, the production cross sections of stop quarks scale as
\begin{equation}
    \sigma_{\tilde{t}_1} \sim \lambda'^{2}_{131} d(x)\cos^2\theta_{\tilde{t}}\;\;\; \mbox{and}\;\;\;
    \sigma_{\tilde{t}_2} \sim \lambda'^{2}_{131} d(x)\sin^2\theta_{\tilde{t}},
\end{equation}
where $d(x)$ is the probability of finding a $d$ quark in the proton with a momentum fraction $x=M_{\tilde{t}_{1,2}}^{2}/s$, where $M_{\tilde{t}_{1,2}}$ denotes the stop masses.
The lighter state does not necessarily have the larger production cross section.
However, in the SUSY parameter space investigated in this paper, $M_{\tilde{t}_{2}}$ is large enough to ensure that the resonant production of $\tilde{t}_{2}$ can be neglected.
Therefore in the following the notation $\tilde t$ will indicate the
lighter $\tilde t_1$.

\begin{figure}[p]
  \begin{center}
    \includegraphics[width=0.62\textwidth]{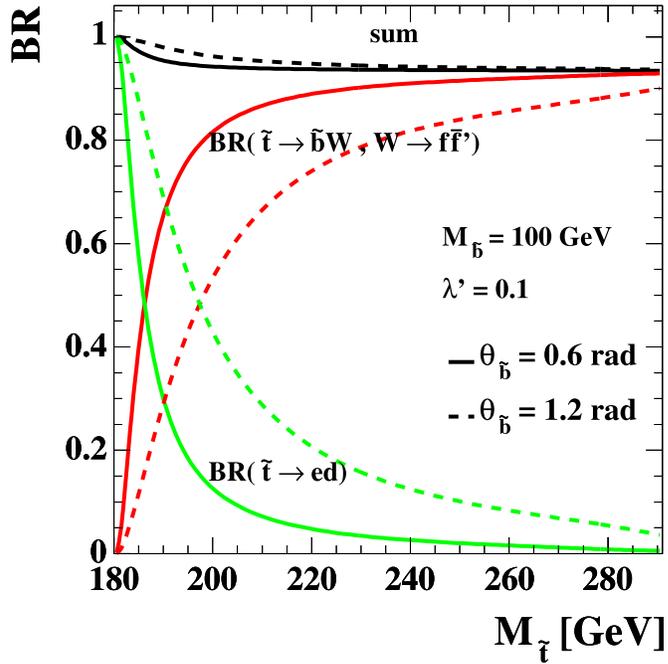}
  \end{center}
  \caption{Examples of the stop branching ratios  as a function of the stop mass for \mbox{$M_{\tilde{b}}=100\gev$} and $\lambda'_{131}=0.1$, when the  fermionic decay modes of the stop via the usual gauge couplings are kinematically suppressed. 
    The solid lines show the branching ratios for $\theta_{\tilde{b}}=0.6$ and the dashed lines for $\theta_{\tilde{b}}=1.2$.
    The sum of the branching ratios is slightly less than one since hadronic $\tau$ decays following  $W \rightarrow \nu \tau$ are not considered.}
  \label{fig:br}
\end{figure}

Searches for fermionic squark decays via their usual gauge couplings (into neutralinos, charginos or gluinos) are presented in \cite{haller}. 
In the present, complementary analysis the SUSY parameter space is chosen such that these decays are kinematically forbidden.
It is moreover assumed that the sbottom quark $\tilde{b}_1$ (denoted by $\tilde{b}$) is lighter than the lightest stop, such that the only possible decay modes are $\tilde{t}\rightarrow \tilde{b} W$ with $W \rightarrow f\bar{f}'$ and the \Rp\hspace{0.075cm} decay into SM fermions, \mbox{$\tilde t \rightarrow e^+ d$}. 
It has been checked that the three--body decays via an off--shell $W$ can be neglected compared with the \Rp\hspace{0.075cm} stop decay for the values of $\lambda'_{131}$ which can currently be probed at HERA.
Thus, only the region $M_{\tilde t}>M_{\tilde b}+M_W$ is investigated here, where the stop quark can decay into a sbottom quark and a real $W$.
The branching ratio $BR_{\tilde{t}\rightarrow \tilde{b} W}$ for this decay mode depends only on the masses of the  squarks involved,  the \Rp\hspace{0.075cm} coupling $\lambda'_{131}$ and  the mixing angle $\theta_{\tilde b}$.
It is proportional to $\cos^2\theta_{\tilde b}$.
This branching ratio is shown for example values of $M_{\tilde b}$ and $\lambda'_{131}$ as a function of the stop mass in figure \ref{fig:br}.
Under the assumption that squark gauge decays into fermions are  kinematically suppressed, the sbottom will subsequently undergo the \Rp\hspace{0.075cm} decay $\tilde b \rightarrow \bar\nu_e d$ and several final states can be investigated depending on the $W$ decay mode.
The four signatures considered in this analysis are given in table \ref{tab:decays}, with the corresponding diagrams shown in \mbox{figure \ref{fig:feynman}.} 
Taking into account the LEP lower bound on the sbottom mass \cite{delphi}, the mass range chosen is \mbox{$180\gev < M_{\tilde t}< 290\gev$} and \mbox{$100\gev < M_{\tilde b}< 210\gev$.}

The interpretation of the results is performed within a Minimal Supersymmetric Standard Model (MSSM) in which the masses of the neutralinos, charginos and gluinos, as well as the couplings between any two SUSY particles and a standard model fermion/boson, are determined by the usual MSSM parameters:
the ``mass'' term $\mu$ which mixes the Higgs superfields, the soft SUSY--breaking mass parameters $M_1$, $M_2$ and $M_3$ for $U(1)$, $SU(2)$ and $SU(3)$ gauginos and $\tan\beta$, the ratio of the vacuum expectation values of the two neutral scalar Higgs fields. 
The usual GUT relations between $M_1$, $M_2$ and $M_3$ are assumed to hold~\cite{SUSY}.
These parameters are defined at the electroweak scale.
All sfermion masses are free parameters in this model, as well as the squark mixings $\theta_{\tilde t}$ and $\theta_{\tilde b}$ and the soft SUSY--breaking trilinear couplings $A_t$ and $A_b$.
The squark mixing parameters, masses and the SUSY parameters $A_t$, $A_b$, $\tan\beta$ and $\mu$ are related by
\begin{eqnarray}
    A_{t} =\frac{M_{\tilde{t}_1}^2-M_{\tilde{t}_2}^2}{2M_t}\cdot\sin 2\theta_{\tilde t}+\mu \cot\beta \;\;\;\mbox{and}\;\;\;
    A_{b} =\frac{M_{\tilde{b}_1}^2-M_{\tilde{b}_2}^2}{2M_b}\cdot\sin 2\theta_{\tilde b}+\mu \tan\beta \ ,
    \label{eq:atab}
\end{eqnarray}
with $M_t$ and $M_b$ being the top and bottom masses, respectively.

\section{The H1 detector}

A detailed description of the H1 detector can be found in \cite{Abt:h1}. 
The H1 detector components relevant to the present analysis are briefly described here. 
The right--handed coordinate system used is centered on the nominal interaction point with the positive $z$--direction defined to be along the incident proton beam.
The Liquid Argon (LAr) calorimeter is used to identify jets and electrons and covers the polar angle range $4^\circ<\theta<154^\circ$ with full azimuthal acceptance.
It has an energy resolution of $\sigma(E)/E \approx 12\%/ \sqrt{E/\gev}\oplus 1\%$ for electrons and \mbox{$\sigma(E)/E \approx 50\%/ \sqrt{E/\gev}\oplus 2\%$} for hadrons, as obtained in test beam measurements.  
The energy measurement is complemented by a calorimeter in the backward region \cite{Abt:h1,19gen}.
The central and forward tracking detectors are used to measure charged particle trajectories, to reconstruct the interaction vertex and to supplement the measurement of the hadronic energy.
The central part of the detector is surrounded by a superconducting magnetic coil with a strength of \mbox{$1.15$ T}.
The iron return yoke is the outermost part of the detector and is
equipped with streamer tubes to form the central muon detector
($4^\circ<\theta<171^\circ$). 
It is supplemented by the forward muon spectrometer
($3^\circ<\theta<17^\circ$) which uses a toroidal magnetic field.
The luminosity is determined from the rate of the Bethe--Heitler process $ep\rightarrow ep\gamma$, detected in a calorimeter located downstream of the interaction point.
The main triggers for the processes investigated are provided by the LAr calorimeter and their efficiencies are close to~$100\%$.

\section{Monte Carlo event generation and simulation}
\label{chap:mc}

For each possible SM background source a detailed Monte Carlo simulation of the H1 detector response is performed.
All processes are generated with an integrated luminosity much higher than that of the data.

To determine the contribution of neutral current (NC) deep inelastic scattering (DIS) events $ep \rightarrow e j X$, where $j$ indicates a jet, the RAPGAP \cite{Jung:1995} event generator is used, which includes the Born, QCD Compton and boson gluon fusion matrix elements. 
Higher order QCD radiative corrections are modelled using leading logarithmic parton showers\cite{Sjostrand:1985xi}.
An important SM background for the bosonic stop decay channels is charged current (CC) deep--inelastic scattering, which is simulated using DJANGO \cite{Schuler:yg}. 
QCD radiation is implemented  to first order via matrix elements, while higher orders are modelled by parton shower cascades generated using the colour--dipole model, as implemented in ARIADNE \cite{Lonnblad:1992tz}. 
In both generators QED radiative effects arising from real photon emission are simulated using HERACLES \cite{Kwiatkowski:1990es}. 
For the simulation of the direct and resolved photoproduction of jets, $ep \rightarrow (e)jj X$, the PYTHIA~6.1 program \cite{Sjostrand:2000wi} is used, which includes light and heavy quark flavours.
It contains the QCD Compton and boson gluon fusion matrix elements and radiative QED corrections.
In the above event generators the parton densities in the proton are taken from the CTEQ5L \cite{Lai:2000} parameterisation. 
The most important SM background to the leptonic $W$ decay channels is the production of $W$ bosons, calculated in leading order (LO) using EPVEC \cite{Baur:1991pp}.
By reweighting the events as a function of the transverse momentum and rapidity of the $W$ boson, next--to--leading order QCD corrections  are accounted for \cite{Diener:2002if}. 
The production of multi--lepton events may also contribute to the SM background for the leptonic $W$ decay channels when one lepton is undetected and some fake missing transverse momentum is reconstructed. 
This process is generated with the GRAPE \cite{Abe:2000cv} program. 

The predictions of the RAPGAP, DJANGO and PYTHIA models are scaled by a factor of $1.2$ for cases where three jets are required. 
This factor accounts for deficiencies in the parton shower model for multi--jet production and is obtained by comparison with data \cite{Caron:2002yc}.


For the SUSY signal simulation and the calculation of its cross section SUSYGEN \cite{susygen} is used which relies on the LO amplitudes for $ed\rightarrow \tilde b W$  given in \cite{stop}.
The parton densities are taken from the CTEQ5L parameterisation and evaluated at the scale of the stop mass.
All bosonic stop decay topologies are simulated for a wide range of stop and sbottom masses in a grid with steps of typically $20\gev$;
for the \Rp\hspace{0.075cm} stop decay only the stop mass is varied.
The events are passed through a detailed simulation of the H1 detector. 
These simulations allow the signal detection
efficiencies as a function of the stop (and sbottom) masses to be determined in the
entire phase space since the mass steps are sufficiently small for linear interpolations to be used.
The variation of the efficiencies with the coupling $\lambda'_{131}$ when the stop mass and width are both large is also taken into account.

\section{Event selection and analysis}

The selection of the event topologies, as given in table~\ref{tab:decays}, relies on the identification of jets, leptons and missing transverse momentum, as detailed below.
The primary interaction vertex has to be reconstructed within $35\cm$ in $z$ of the nominal position of the vertex.
Non--$ep$ background is rejected by searching for event topologies typical of cosmic ray and beam--induced background \cite{NCCCpaper} and the event timing is required to be consistent with the $ep$ bunch crossing.

\subsection{Particle identification}

The {\bf electron} identification is based on the measurement of a compact and isolated electromagnetic shower in the LAr calorimeter.
The hadronic energy within a cone defined by $R=\sqrt{(\Delta \eta)^2+(\Delta \phi)^2}<0.5$ around the electron direction is required to be below $2.5\%$ of the electron energy. 
Here, $\eta = -\ln(\tan(\theta/2))$ is the pseudorapidity and $\phi$ denotes the azimuthal angle.
For electron polar angles in the region $10^\circ < \theta_e <  140^\circ$ a high quality track is also required to be associated to the electromagnetic cluster.
This allows efficient rejection of photons which convert late in the central tracker material.

The {\bf muon} identification is based on the measurement of a track segment or an energy deposit in the instrumented iron associated with a charged particle track in the inner tracking systems \cite{Andreev:2003pm}.
In addition, a track in the forward muon system is taken as a muon candidate.
The muon momentum is measured from the track curvature in the solenoidal or toroidal magnetic field.
A muon candidate should not deposit more than $8\gev$ in the LAr calorimeter.
The distance between the muon candidate and the nearest track is required to be $R>0.5$.
For muon pairs a cut on the track opening angle and polar angle sum is applied to reject cosmic muons.

{\bf Jets} are reconstructed from energy deposits in the LAr calorimeter combined with well measured tracks using a modified inclusive $k_{\bot}$ algorithm \cite{Ellis:1993tq,Catani:1993hr} in the laboratory frame.
Electron and muon candidates are excluded from the algorithm.
Only jets in the polar angle range $7^\circ < \theta_{Jet} < 140^\circ$ are considered to ensure that they are reliably measured.
To reject electrons which are misidentified as jets, topological criteria for electron--jet separation are applied. 
About $80\%$ of  fake jets and $3\%$ of  genuine jets are rejected, as determined from simulations.

The {\bf missing transverse momentum} \ptmiss is derived from a summation over all identified particles and energy deposits in the event.
In the channels where one or more neutrinos are expected, an event is
only accepted if the energy and the momentum in the beam direction
fulfil \mbox{$\sum_i (E_i-P_{z,i})<50\gev$}, where $E_i$ is the energy
and $P_{z,i}$ is the $z$ component of the momentum and the index $i$
runs over all hadronic and electromagnetic objects and muons. 
This requirement reduces the contamination due to badly measured NC
 DIS events\footnote{A NC DIS event is expected to have 
   $\sum_i (E_i-P_{z,i}) = 2 E_e = 55.2\gev$ due to energy and 
   momentum conservation.}
 where fake missing transverse momentum is reconstructed.

\subsection{Systematic uncertainties}
\label{chapt:syst}

The sources of experimental and theoretical systematic uncertainties considered in this analysis are described in the following.
They are added in quadrature.
\begin{itemize}
\item  
The electromagnetic energy scale uncertainty is between $0.7\%$ and $3\%$ depending on the particle's impact position on the LAr calorimeter  \cite{NCCCpaper}. 
The  uncertainty on the polar angle of electromagnetic clusters  varies between $1$~mrad and $3$~mrad, depending on $\theta$ \cite{NCCCpaper}.
The uncertainty on the azimuthal angle is $1$ mrad.
The tracking efficiency is known with a precision of $2\%$ for polar angles above $37^\circ$ and deteriorates to $15\%$ in the forward region.

\item The muon $P_T$ scale uncertainty is  $5\%$. 
  The uncertainty on the polar angle is $3$ mrad and on the azimuthal angle is $1$ mrad.

\item The hadronic energy scale of the LAr calorimeter is known to $2\%$. 
  The uncertainty on the jet polar angle determination is $5$ mrad for $\theta<30^\circ$ and $10$ mrad for
  $\theta>30^\circ$.
  
\item The uncertainty on the integrated luminosity results in an overall normalisation error of $1.5\%$.
  
\item Depending on the SM production process different theoretical uncertainties are used.
  These amount to $15\%$ for $W$ production, $10\%$ for NC DIS processes and $15\%$ for photoproduction.
  For $ep \rightarrow \nu j j j X$ reactions, the theoretical uncertainties are about $20\%$, which takes into account the deficiencies of the parton shower modelling of multi-jet production.
  
\item For the SUSY signal detection efficiencies, an uncertainty of $10\%$ is assumed resulting mainly from the linear interpolation in the grid of simulated mass values.
  An additional theoretical systematic error, mainly due to the uncertainties on the parton densities, affects the signal cross section. 
  This uncertainty varies between $12\%$ at lower stop masses
  (\mbox{$x\approx 0.3$}) up to $50\%$ for stop masses of $290\gev$
  ($x\approx 0.8$) at \mbox{$\sqrt{s}=319\gev$}.
  An additional uncertainty of $7\%$ on the signal cross section
\cite{haller} arises from the variation of the scale at which the
parton densities are evaluated.
\end{itemize}

\subsection{Analysis of the bosonic stop decay channels}

According to table \ref{tab:decays} the bosonic stop decay leads to three different final state topologies.
If the $W$ boson decays into leptons, the signature is a jet, a lepton (electron or muon) and missing transverse momentum ($je$\ptmiss channel and $j\mu$\ptmiss channel).
The $W$ decay into $\nu_{\tau}\tau$, where \mbox{$\tau\rightarrow\mbox{hadrons}+\nu$}, is not investigated in this paper.
If the $W$ decays into hadrons the signature is three jets and missing transverse momentum ($jjj\hspace{-0.2cm}$ \ptmiss channel). 
The selection of the final states analysed is described in the following sections.

\mathversion{bold}
\subsubsection{The channels $\tilde t \rightarrow je$\ptmiss and $\tilde t \rightarrow j\mu$\ptmiss} 
\mathversion{normal}
The selection criteria for the $je$\ptmiss and $j\mu$\ptmiss channels are the following.
\begin{itemize}
\item A lepton must be found with $P_T^\ell > 10 \gev$ and with polar angle $5^\circ < \theta_{e} < 120^\circ$ for the electron and $10^\circ < \theta_{\mu} < 120^\circ$ for the muon.
\item A jet must be found with $P_T^{Jet}>10\gev$ within the angular range $7^\circ < \theta_{Jet} < 140^\circ$.
\item The total missing transverse momentum must satisfy \ptmiss $> 12 \gev$.
\item The difference in azimuthal angle between the lepton $l$ and the direction of the system $X_{tot}$, composed of all other measured particles in the event, must be \mbox{$\Delta \phi(l-X_{tot})<165^\circ$}.
  NC background events with a mismeasured electron are rejected by this cut.
\item The azimuthal balance of the event must satisfy $V_{AP}/V_P<0.3$, where $V_{AP}/V_P$ is defined as the ratio of the anti--parallel component to the parallel component of the measured calorimetric transverse momentum with respect to the direction of the total calorimetric transverse momentum \cite{vapvp}. 
  This cut removes NC DIS events which generally have high values of $V_{AP}/V_P$.
\item In the $je$\ptmiss channel, the variable $y_e = 1 - E'_e(1-\cos\theta_e)/(2E_e)$, where $E'_e$ denotes the energy of the scattered electron, is required to fulfil $y_e>0.3$.
  This cut reduces the remaining NC DIS background, since particles coming from a bosonic stop decay will be boosted in the forward direction, leading to a rising $d\sigma/dy$ distribution.
  This contrasts with the steeply falling $d\sigma/dy \sim y^{-2}$ distribution of NC DIS events.
\end{itemize}

\begin{figure}[p]
  \begin{center}
    \includegraphics[width=1\textwidth]{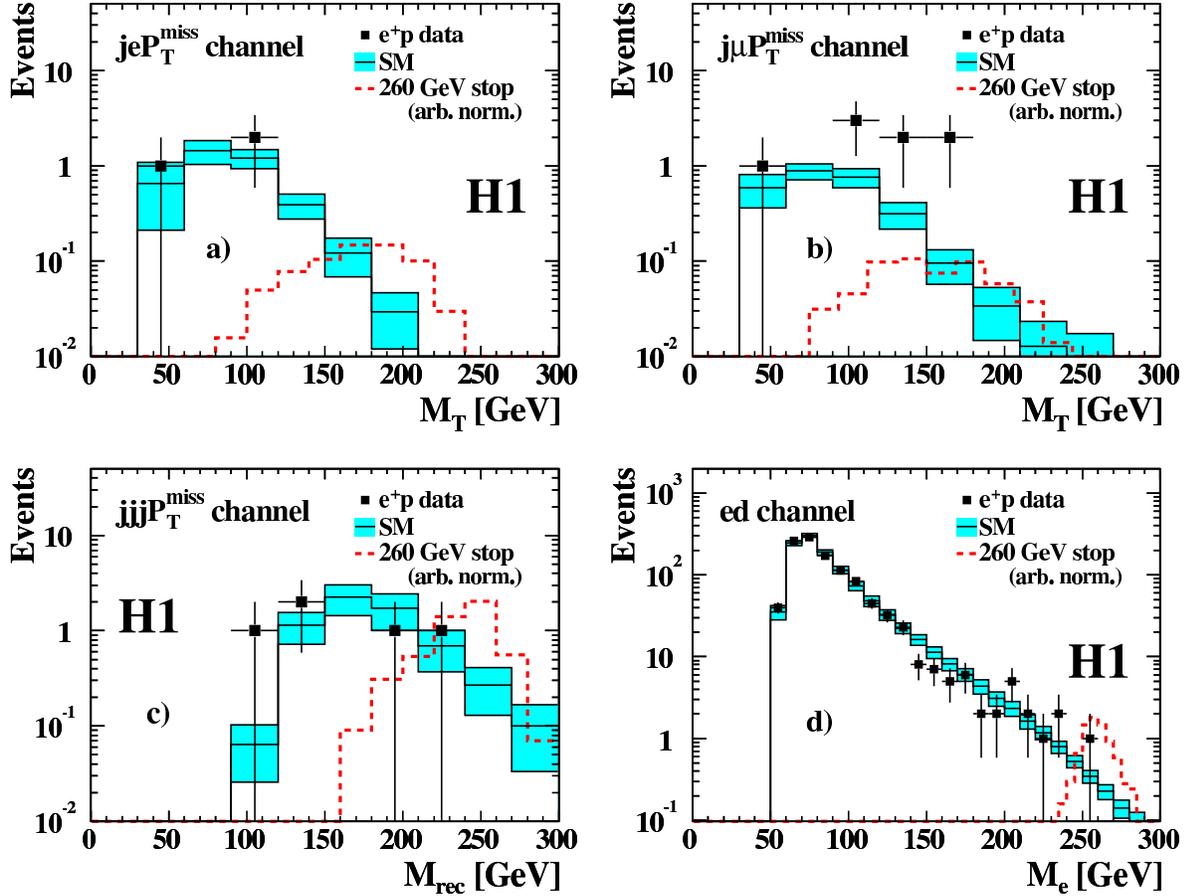}
  \end{center}
  \vspace{-0.5cm}
  \caption{Mass spectra for the H1 $e^+p$ data: a) transverse mass of the $je$\ptmiss channel; b) transverse mass of the $j\mu$\ptmiss channel; c) reconstructed mass of the $jjj$\ptmiss channel; d) invariant mass distribution of the $ed$ channel. The data are compared with the SM expectations with the systematic uncertainties shown as the shaded band. The expected signal from a $\tilde t$ with mass $260\gev$ is also shown with arbitrary normalisation.}
  \label{fig:mass}
\end{figure}

The stop mass cannot be reconstructed in these channels since there are two neutrinos in the final state. 
Therefore the transverse mass distributions are investigated.
The transverse mass is defined as 
\begin{equation}
  M_T = \sqrt{(\mbox{\ptmiss} +P_T^\ell+P_T^{Jet})^2 - (\vec{\mbox{\ptmiss}}+\vec{P}_T^\ell+\vec{P}_T^{Jet})^2},
\end{equation}
where $\vec{\mbox{\ptmiss}}$, $\vec{P}_T^\ell$ and $\vec{P}_T^{Jet}$ are the missing transverse momentum, lepton and jet momentum, respectively.
The transverse mass distributions are shown in figures \ref{fig:mass}a and \ref{fig:mass}b.
In the $je$\ptmiss channel, $3$ events are found while the expectation from the SM is $3.84 \pm 0.92$ events.
In the $j\mu$\ptmiss channel, $8$ events are found while $2.69 \pm 0.47$ are expected.
This slight excess could be interpreted as a scalar top decaying via $\tilde t \rightarrow \tilde b W$ \cite{stop}. 
All $11$ events in the \mbox{$je$\ptmiss} and \mbox{$j\mu$\ptmiss} channels were also found in \cite{Andreev:2003pm}, 
where additional events were selected since there were no explicit jet requirements.
Between $60\%$ and $70\%$ of the SM expectation arises from the production of real $W$ bosons. 
The numbers of events and the SM expectations can be found in table \ref{tab:evex}. 
The stop signal efficiency amounts to  typically $35\%$--$45\%$ for the \mbox{$je$\ptmiss} channel and  $30\%$--$40\%$ for the \mbox{$j\mu$\ptmiss} channel and depends mainly on $M_{\tilde{t}}$ and $M_{\tilde{b}}$.

\mathversion{bold}
\subsubsection{The channel $\tilde t \rightarrow jjj$\ptmiss} 
\mathversion{normal}

For the $jjj$\ptmiss final state topology the following criteria are required.
\begin{itemize}
\item Three jets must be found with $P_T^{Jet 1} > 20 \gev$, $P_{T}^{Jet 2} > 15 \gev$ and $P_{T}^{Jet 3} > 10 \gev$, each with polar angle $7^\circ < \theta_{Jet} < 140^\circ$. 
\item The total missing transverse momentum must satisfy \ptmiss$ > 25 \gev$.
\item The selection is restricted to $y_{h}>0.4$ exploiting the different $y_{h}$ distributions of the stop signal and the SM background.
  Here, $y_{h}$ is calculated using \mbox{$y_{h}=\sum_h (E_h-P_{z,h})/2E_e$} \cite{JB}, where $\sum_h (E_h-P_{z,h})$ is calculated from the hadronic energy deposited in the detector.
\end{itemize}

Assuming that only one neutrino is present in the event and applying the constraints $\vec{\mbox{\ptmiss}}=\vec{P}_T^\nu$ and \mbox{$\sum_i (E_i-P_{z,i}) + (E_\nu-P_{z,\nu}) = 2E_e$}, the neutrino four--vector can be calculated. 
Hence, the invariant mass $M_{rec}$ can be reconstructed in this final state topology with a mass resolution of about $15\gev$.
In figure \ref{fig:mass}c the reconstructed mass distribution for the $jjj$\ptmiss channel is shown.
A total of $5$ events are found while $6.24 \pm 1.74$ are expected from SM processes (see table \ref{tab:evex}).
The SM background arises predominantly from CC DIS processes. 
The stop detection efficiency is typically $35\%$--$50\%$ in this final state topology.

\mathversion{bold}
\subsection{Analysis of the R--parity violating stop decay channel $\tilde t \rightarrow ed$}
\mathversion{normal}

For stop and sbottom masses for which $M_{\tilde{t}}\approx M_{\tilde{b}}+M_W$, the \Rp\hspace{0.075cm} decay $\tilde t \rightarrow ed$ becomes dominant (see figure \ref{fig:br}).
Events from this process are characterised by high $Q^2$ NC DIS--like topologies.
The momentum transfer squared, obtained from the scattered electron, is defined by $Q_e^2=(P_T^e)^2/(1-y_e)$. 
Both the stop decay and the NC DIS final states consist of a jet and an electron with high transverse momenta.
However, the distributions of the events in mass $M_e= \sqrt{x_e s}$ and $y_e$ are different. 
Here, the Bjorken variable $x_e$ is related to the other kinematic quantities by $Q_e^2 = x_ey_e s$. 
Stop decays via \Rp\hspace{0.075cm} lead to a resonance in the $M_e$ distribution.
In addition, stop quarks decay isotropically in their rest frame leading to a flat $d\sigma/dy$ distribution, contrasting with that of NC DIS events. 

The selection criteria for the $\tilde t \rightarrow ed$ channel are the following.
\begin{itemize}
  \item The longitudinal momentum loss is limited by requiring
    $40\gev<\sum_i (E_i-P_{z,i})<70\gev$.
  \item An electron must be found with $P_T^e>20\gev$ and with polar angle $5^\circ < \theta_{e} < 120^\circ$.
  \item A jet must be found with $P_T^{Jet}>20\gev$ and with polar angle $7^\circ < \theta_{Jet} < 140^\circ$.
  \item The total missing transverse momentum and $\sqrt{P_{T}^e}$ must fulfil \ptmiss$/\sqrt{P_{T}^e}<4\sqrt{\gev}$, which takes into account the energy resolution of the LAr calorimeter.
  \item Only events with $Q_e^2>2500\gev^2$ are considered.
  \item The selection is restricted to $y_e<0.9$ to avoid the region where migration effects due to QED radiation in the initial state are largest.
    Background from photoproduction, where a jet is misidentified as an electron, is also suppressed by this cut.
  \item In order to maximise the signal sensitivity, a mass dependent lower $y_e$ cut is applied as in \cite{haller}, which exploits the differences in the $M_e$ and $y_e$ distributions between the SUSY signal and the DIS background.
\end{itemize}

The $M_e$ spectrum for data and for the SM expectation are shown in figure \ref{fig:mass}d for all H1 $e^+p$ data.
The resolution in $M_e$ is between $5\gev$ and $9\gev$, depending on the stop mass.
No significant deviation from the SM is found. 
In particular, at masses above $\sim 180\gev$ where the stop signal is searched for, no significant peak is observed in the data. 
A total of $1100$ events are found, while  $1120 \pm 131$ are expected from SM processes, mainly from NC DIS events. 
The numbers of events and the SM expectations can be found in table \ref{tab:evex}.
In the $ed$ channel, the typical stop signal efficiency is about $30\%$--$45\%$.

\section{Results of SUSY analysis}
 
\subsection{Interpretation of bosonic stop decay searches}
\label{interpretation}
In the $j\mu$\ptmiss channel a slight excess of events compared with the SM expectation is observed, confirming the previous H1 analysis \cite{Andreev:2003pm}.
All other channels are in good agreement with the SM (see table \ref{tab:evex}).

\begin{figure}[p]
  \begin{center}
    \includegraphics[width=0.5\textwidth]{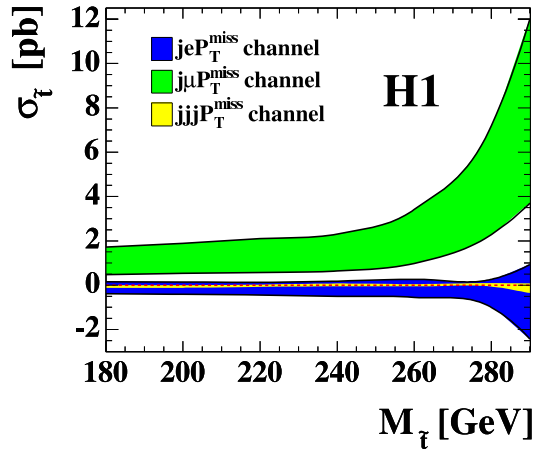}
  \end{center}
  \vspace{-0.8cm}
  \caption{Bands representing the allowed stop cross section regions $\sigma_{\tilde{t}}\pm \Delta\sigma_{\tilde{t}}$ as a function of the stop mass as obtained from the analysis of each bosonic stop decay channel.}
  \label{fig:band}
\end{figure}

Assuming the presence of a stop of mass $M_{\tilde{t}}$ decaying bosonically, the observed
event yields are used to determine the allowed range for a
stop production cross section $\sigma_{\tilde t}$.
The number of observed and expected events satisfying the relevant
selection cuts, $N_{data}$ and $N_{SM}$, are integrated within a mass
bin (transverse mass bin) around the calculated stop mass (transverse
mass), corresponding to the decay channel under consideration.
The width of the mass bin is adjusted to the expected mass resolution, such that each bin contains events reconstructed within $\pm 2$ standard deviations of the given stop mass.
A signal cross section $\sigma_{\tilde t}$ dependent on the stop mass 
can be determined from $N_{data}$ and $N_{SM}$ in each bosonic decay channel by folding in the signal efficiency $\epsilon$, the $\tilde t$ and $W$ branching ratios \mbox{$BR_{\tilde t \rightarrow \tilde b W}\cdot  BR_{W\rightarrow f \bar{f}'}$} 
and taking into account the integrated luminosities $\mathcal{L}_{301}$ and $\mathcal{L}_{319}$:
\begin{equation}
\sigma_{\tilde t}(M_{\tilde{t}})=\frac{N_{data}-N_{SM}}{\epsilon \cdot BR_{\tilde t \rightarrow \tilde b W}\cdot  BR_{W\rightarrow f \bar{f}'}}\cdot \frac{1}{r_{\sigma}\cdot \mathcal{L}_{301} + \mathcal{L}_{319}}.
\end{equation}
Here, $r_{\sigma}$ is the ratio of the theoretical stop production cross sections at $\sqrt{s} = 301\gev$ and $\sqrt{s} = 319\gev$.
The branching ratio for $\tilde t \rightarrow \tilde b W$ is assumed to be $BR_{\tilde t \rightarrow \tilde b W}=100\%$.
The uncertainty on the cross section, $\Delta\sigma_{\tilde t}$, is
determined from the statistical error on the number of observed events
and the systematic uncertainty on the SM prediction.
The bands in figure \ref{fig:band} represent the allowed cross section regions 
for all bosonic decay channels. The band for the $jjj$\ptmiss channel is narrow due to the large branching ratio $BR_{W\rightarrow q \bar{q}'}$.

From figure \ref{fig:band} it can be seen that the stop interpretation of the excess seen in the $j\mu$\ptmiss channel is not supported by the other decay modes. 
For instance, the probability that the observed event rate in the $jjj$\ptmiss channel fluctuates upwards to produce at least the number of events expected on the basis of the signal in the \mbox{$j\mu$\ptmiss} channel is around $1\%$, depending slightly on the stop mass.
Hence, exclusion limits on the \Rp\hspace{0.075cm} SUSY model described in Section \ref{chap:phen} are derived.

\subsection{Exclusion limits in the MSSM}
\label{chap:limits}

The results from the selection channels considered in this paper are combined to derive constraints in the MSSM. 
For a given set of parameters, the full supersymmetric mass spectrum and the branching ratios of all stop and sbottom decay modes are calculated.
An upper limit $\sigma_{lim}$ on the stop production cross section is calculated at the $95\%$ confidence level (CL) using a modified frequentist approach based on likelihood ratios \cite{junk}.

Each considered channel contributes via its branching ratio,
the signal efficiencies and the number of observed and expected events
within  sliding mass bins (transverse mass bins). 
Although the selection criteria for the various channels are not
explicitly exclusive, it was checked that double counting of  events is negligible. 
The given set of model parameters is excluded if it predicts a cross section which is larger than $\sigma_{lim}$.

In order to investigate systematically the dependence of the sensitivity on the MSSM para\-meters, a scan of the SUSY parameter space is performed.
The SUSY parameter space is selected such that the combined branching ratio is
\begin{equation}
  BR_{tot} = BR_{\tilde t \rightarrow ed} + BR_{\tilde t \rightarrow \tilde b W}\cdot BR_{\tilde b \rightarrow \nu_e d}>85\%.
\end{equation}
The parameter $M_2$ is set to $1000\gev$ and $\mu$ is restricted to the range $400\gev < \mu < 1000\gev$, which ensures that the gaugino masses are large.
The mixing angles $\theta_{\tilde t}$ and $\theta_{\tilde b}$ are allowed to vary between $0.6$ rad and $1.2$ rad.
For given values of $\tan\beta$, $A_t$ and $A_b$, the parameters $M_{\tilde{t}_1}, M_{\tilde{b}_1}, \theta_{\tilde t}, \theta_{\tilde b}$ and $\mu$ are scanned.
Here, $A_t$ and $A_b$ are only needed to determine the masses of the heavier stop and sbottom, according to equation (\ref{eq:atab}), they are set to $A_t = A_b = -100\gev$.

\begin{figure}[p]
  \begin{center}
    \includegraphics[width=1\textwidth]{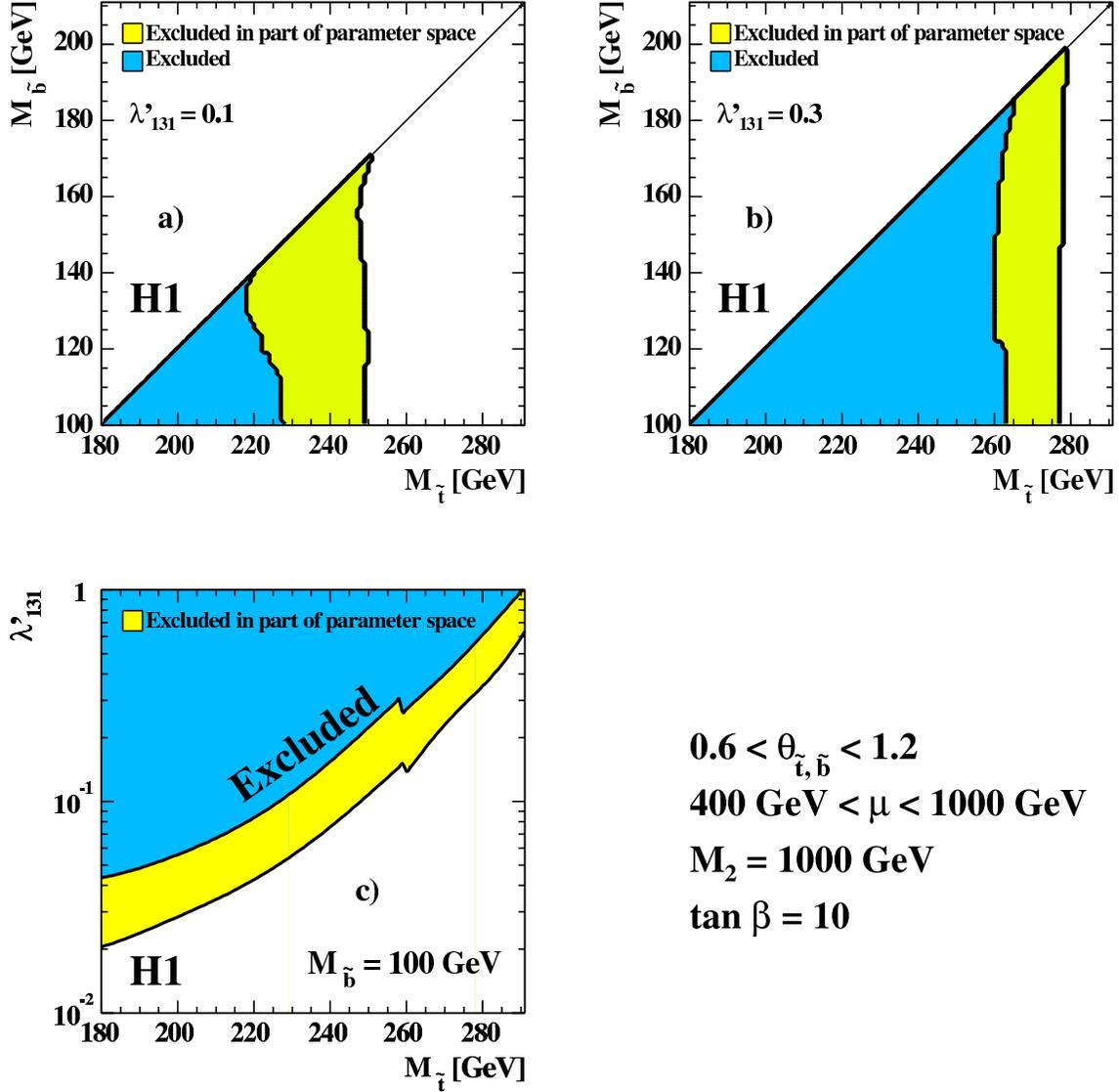}
  \end{center}
  \caption{Exclusion limits at the $95\%$ CL in the $(M_{\tilde t}, M_{\tilde b})$ plane for a) $\lambda'_{131}=0.1$ and b) $\lambda'_{131}=0.3$. 
    c) Exclusion limits at the $95\%$ CL on the \Rp\hspace{0.075cm} coupling $\lambda'_{131}$ as a function of the stop mass for $M_{\tilde b}=100\gev$.
    The limits are derived from a scan of the MSSM parameter space as indicated in the legend. 
    The two full curves indicate the regions excluded in all (dark) or part (light) of the parameter space investigated.}
  \label{fig:limit}
\end{figure}

For each point in the 5--dimensional parameter space an upper bound on the coupling $\lambda'_{131}$ is obtained.
The resulting limits are given for two cases: (i) every point of the scanned SUSY parameter space is excluded, (ii) at least one point in the scanned SUSY parameter space is excluded.
The resulting limits obtained for $\tan\beta=10$ are shown in figure \ref{fig:limit}a and \ref{fig:limit}b in the \mbox{$(M_{\tilde t}, M_{\tilde b})$ plane} for $\lambda_{131}'=0.1$ and $\lambda_{131}' = \sqrt{4\pi\alpha_{em}}=0.3$.
At $\lambda_{131}'=0.1$ stop masses $M_{\tilde t}\lsim 250\gev$ can be excluded, while masses $M_{\tilde t}\lsim 275\gev$ are excluded at a Yukawa coupling of electromagnetic strength, i.e. $\lambda'_{131}=0.3$. 
The resulting limits projected on the $(M_{\tilde t},\lambda'_{131})$ plane for \mbox{$M_{\tilde b}=100\gev$} are shown in figure \ref{fig:limit}c. 
For $M_{\tilde t}=200\gev$,  couplings $\lambda'_{131}\gsim 0.03$ are ruled out and for $M_{\tilde t}=275\gev$ the allowed domain is $\lambda'_{131} \lsim 0.3$.
The limits do not significantly depend on $\tan\beta$ or on $M_2$, provided that $M_2>400\gev$, which has been checked by repeating the analysis with $\tan\beta=2$ or $M_2=400\gev$.


\section{Conclusions}

A search is performed for scalar top quarks resonantly produced in $e^+p$ collisions at HERA in R--parity violating SUSY models. 
Final state topologies resulting from R--parity conserving bosonic stop decays or R--parity violating direct decays are considered.
In the $j\mu$\ptmiss channel, a slight excess of events compared with the SM expectation is observed. 
Nevertheless, no evidence for stop production is found, since this excess is not supported by the other three channels analysed in the present paper.
 
For the first time, direct constraints on stop quarks decaying bosonically are derived.
Including the direct \Rp\hspace{0.075cm} stop decay, mass dependent limits on the coupling $\lambda_{131}'$ are obtained within 
a Minimal Supersymmetric Standard Model.
In a large part of the model parameter space, the existence of stop quarks coupling to an $e^+d$ pair with masses up to $275 \gev$ is excluded at the $95 \%$ CL for a strength of the Yukawa coupling of $\lambda_{131}' = \sqrt{4\pi\alpha_{em}}=0.3$.

\section*{Acknowledgements}

We are grateful to the HERA machine group whose outstanding efforts have made this experiment possible. 
We thank the engineers and technicians for their work in constructing and maintaining the H1 detector, our funding agencies for financial support, the DESY technical staff for continual assistance and the DESY directorate for support and for the hospitality which they extend to the non DESY members of the collaboration.

\newpage

\begin{table}[p]
  \begin{center}
    \begin{tabular}{|c|cccc|c|}
      \hline
      & & & & &\\[-2ex]
      {\bf Channel} & \multicolumn{4}{c|}{{\bf Decay process}} & {\bf Signature}\\
      & & & & &\\[-2ex]
      \hline
      \hline
      & &  & & &\\[-2ex]
      & $\tilde{t} \rightarrow \tilde{b}\;W$ & & & &\\
      & & \hspace{-1.2cm}$\stackrel{\lambda'}\hookrightarrow d\bar{\nu}_e$ & & &\\
      & &  & & &\\[-2ex]
      $je$\ptmiss & & & \hspace{-1.4cm}$W \rightarrow e{\nu}_e$ & & jet + $e$ + \ptmiss\\
      & & & & \hspace{-1.8cm}{\small $\rightarrow \tau{\nu}_\tau\rightarrow e\nu\nu\nu$} & \\
      $j\mu$\ptmiss & & & \hspace{-1.3cm}$W \rightarrow \mu{\nu}_{\mu}$ & &  jet + $\mu$ + \ptmiss\\
      & & & & \hspace{-1.8cm}{\small $\rightarrow \tau{\nu}_\tau \rightarrow \mu\nu\nu\nu$} & \\
      $jjj$\ptmiss & & & \hspace{-1.5cm}$W \rightarrow q\bar{q}'$ & & 3 jets + \ptmiss\\
      & &  & & &\\[-2ex]
      \hline
      & &  & & &\\[-2ex]
      $ed$ & $\tilde{t} \stackrel{\lambda'}\rightarrow ed$ && &  & jet + high $P_T$ $e$\\[0.5ex]
      \hline
    \end{tabular}
  \end{center}
  \caption{Analysed stop decay channels  in \Rp\hspace{0.075cm} SUSY.
    The \Rp\hspace{0.075cm} processes are indicated by the coupling $\lambda'$, and \ptmiss denotes the missing transverse momentum.}
  \label{tab:decays}
\end{table}
\begin{table}[p]
  \begin{center}
    \begin{tabular}[t]{|l|c|l|c|l|c|l|}
      \hline
      & \multicolumn{2}{c|}{} & \multicolumn{2}{c|}{} & \multicolumn{2}{c|}{}\\[-2ex]
      {\bf Channel} & \multicolumn{2}{c|}{\mathversion{bold}$\sqrt{s} =301\gev$\mathversion{normal}} & \multicolumn{2}{c|}{\mathversion{bold}$\sqrt{s} =319\gev$\mathversion{normal}} & \multicolumn{2}{c|}{{\bf combined}} \\
      & \multicolumn{2}{c|}{} & \multicolumn{2}{c|}{} & \multicolumn{2}{c|}{}\\[-2ex]
      & {\bf data} & {\bf SM expectation} & {\bf data} & {\bf SM expectation} & {\bf data} & {\bf SM expectation} \\
      & & & & & &\\[-2ex]
      \hline
      \hline 
      & & & & & &\\[-2ex]
      $je$\ptmiss & $1$ & $1.16 \pm 0.28$ & $2$ & $2.68 \pm 0.64$ & $3$ & $3.84 \pm 0.92$ \\
      &  & {\small ($W$: $0.75 \pm 0.12$)} & & {\small ($W$: $1.80 \pm 0.29$)} & & {\small ($W$: $2.55 \pm 0.41$)}\\
      & & & & & &\\[-2ex]
      \hline
      & & & & & &\\[-2ex]
      $j\mu$\ptmiss & $4$ & $0.84 \pm 0.14$ & $4$ & $1.85 \pm 0.33$ & $8$ & $2.69 \pm 0.47$ \\
      &  & {\small ($W$: $0.57 \pm 0.09$)} & & {\small ($W$: $1.36 \pm 0.22$)} & & {\small ($W$: $1.93 \pm 0.31$)} \\
      & & & & & &\\[-2ex]
      \hline
      & & & & & &\\[-2ex]
      $jjj$\ptmiss & $1$ & $1.91 \pm 0.54$ & $4$ & $4.33 \pm 1.21$ & $5$ & $6.24 \pm 1.74$ \\
      & & & & & &\\[-2ex]
      \hline
      & & & & & &\\[-2ex]
      $ed$ & $366$ & $384 \pm 45$ & $734$ & $736 \pm 86$ & $1100$ & $1120 \pm 131$ \\[0.5ex]
      \hline
    \end{tabular}
  \end{center}
  \caption{Total number of selected events in the various stop decay channels for the H1 $e^+p$ data at $\sqrt{s} =301\gev$, $\sqrt{s} =319\gev$ and the combined data set. For the $je$\ptmiss and $j\mu$\ptmiss channels the SM expectations arising from $W$ production are given in brackets.}
  \label{tab:evex}
\end{table}

\newpage

\end{document}